\documentstyle[12pt,epsfig]{article}
\title{\Large\bf  Non-Leptonic Decays of B Mesons and Strong 
Coupling Constants}

\author{\large Gustav Kramer and
  Cai-Dian L\"u\footnote{Alexander von Humboldt Foundation Fellow.
}  \\
{\small II Institut f\"ur Theoretische Physik, Universit\"at Hamburg,
22761 Hamburg, Germany}\thanks{Supported by Bundesministerium f\"ur 
Bildung und
Forschung, Bonn, under Contract 057HH92P(0) and EEC Program ``Human Capital
 and Mobility'' through Network ``Physics at High Energy Colliders'' under
Contract CHRX-CT93-0357 (DG12COMA).}
}
\date{}

\begin{document}
\maketitle

\begin{picture}(0,0)(0,0)
\put(350,250){{\large hep-ph/9707304}}
\put(350,230){{\large DESY-97-134}}
\put(350,210){{\large July 1997}}
\end{picture}

\begin{abstract}
Non-leptonic decays of B mesons into two mesons or meson resonances are 
studied on the basis of two versions of simple pole-dominance models 
involving scalar, vector, pseudoscalar and axial-vector poles. 
The results are compared with those obtained from the usual factorization 
model and used to obtain information on strong coupling constants between 
B meson and one light or one charmed meson, respectively. These coupling
 constants are compared to results from various QCD sum rule calculations.

\end{abstract}

\newpage

\section{Introduction}

 Non-leptonic weak decays of B mesons are very interesting for
 several reasons. 
First, CP violation in the B-meson system will eventually give us 
information about the CP violating phase in the Cabibbo
-Kobayashi -Maskawa (CKM) mixing matrix \cite{1}.
Second, non-leptonic weak decays will give additional clues for 
determining the absolute values of the quark mixing parameters, 
in particular the ratio $|V_{ub} /V_{cb}|$, although it is expected 
that more solid information will come from semi-leptonic B decays. 
Last, the dynamics of the non-leptonic weak decays in the framework of
the standard model is only poorly understood. 
One of the problems in calculating the transition amplitudes for 
non-leptonic weak decays is that one needs to evaluate the hadronic 
matrix elements of certain four-quark operators which can be done 
in QCD only with non-perturbative methods.
The usual route to calculating these hadronic matrix elements is to 
start from the effective, QCD corrected, Hamiltonian for the $\Delta b=1$
 non-leptonic decays in the six-quark model, i.e. including the t quark
\cite{2,3,4}.
This gives the weak Hamiltonian in terms of four-quark operators. 
For computing the hadronic matrix elements of these four-quark operators 
the factorization approximation is used \cite{5,6,7}.
Then the hadronic matrix elements are given in terms of current matrix 
elements (matrix elements of two-quark operators) as they appear also 
in semi-leptonic decays.
These current matrix elements are much easier to calculate and many models  
\cite{8,9,10} have been proposed for them. 
When these current matrix elements are approximated by a single pole based on
the idea of vector dominance we arrive at the pole-dominance model of 
 two-body non-leptonic decays.
This model has been applied to the calculation of non-leptonic decays 
of charmed mesons \cite{11}.
Many years ago, this model was used already for the discussion of 
strange particle decays. 
In particular, it is known, that this model provides a basis for a 
 description of $K\to 2 \pi$ decays \cite{12}.

In this work we apply the pole-dominance model to two-body non-leptonic
 B decays and show its strong relationship to the usual factorization model 
\cite{7,13}.
Therefore this latter model will be our starting point where we also use 
as input the parameters of the current matrix elements as given in \cite{7}. 
This allows us to extract various strong coupling constants between B mesons,
 their vector and axial-vector counterparts, and light and charmed mesons. 
These coupling constants and also the form factors of the current matrix 
elements will be compared with other information coming from QCD sum rule 
and QCD lattice calculations. 
For simplicity, we consider only the lowest lying poles. 
The inclusions of higher lying poles would make the extraction of 
the strong coupling constants impossible.

The outline is as follows: 
In section 2, we explain the relationship of the usual approach with 
the pole-dominance model using as example a collection of B 
decays into two pseudoscalar-mesons ($B \to PP$). 
We calculate the two-body branching ratios in parallel for the usual 
approach and two versions of the pole-dominance framework. 
In one version we calculate the residues of the poles on the mass shell 
of the intermediate state. 
This reproduces the usual approach. 
The other version follows essentially the work of Bedaque et al.\cite{11}
for D decays where a particular off-shell extrapolation has been adopted.
These results are 
compared to experimental data when they are available. In section 3, 
we proceed to the more complicated $B\to VP$ and $B\to VV$ channels,
 where V are light or charmed vector mesons. Section 4 is reserved 
for a summary and some conclusions.

\section{B decays into two pseudoscalar mesons}

To start, we choose as an example for the decays $B \to PP$, the channel 
$\bar B^0 \to \pi^+ \pi^-$.
The relevant effective weak Hamiltonian can be parameterized as 
\begin{equation}
  \label{1}
  {\cal H}_{eff} = \frac{4G_F}{\sqrt{2}} \left [ V_{ub}V_{ud}^*
(C_1O_1 + C_2 O_2)\right] + h.c. ,
\end{equation}
where the $C_i$ (i=1,2) are the short-distance Wilson coefficients 
defined at a scale $\mu$ of the order of the heavy quark mass $m_b$ 
and the $O_i$ are the local quark operators with the appropriate 
quantum numbers. 
For simplicity we have neglected the strong and 
electroweak penguins. The electroweak penguins are found to give 
small contributions to the mode $\bar B^0 \to \pi^+ \pi^-$. 
The strong penguins could be included easily \cite{14}.

The explicit expressions of the operators $O_1$ and $O_2$ are:
\begin{eqnarray}
  \label{2}
  O_1 &=& (\bar d_\alpha \gamma ^\mu L u_\beta )
(\bar u _\beta \gamma_\mu L b_\alpha) ,\\\nonumber
 O_2 &=& (\bar d_\alpha \gamma ^\mu L u_\alpha )
(\bar u _\beta \gamma_\mu L b_\beta),
\end{eqnarray}
where $\alpha, \beta$ are colour indices and $L=\frac{1}{2}(1-\gamma_5)$.
For the Wilson coefficients at the scale $\mu=m_b =4.8$GeV 
we use the values \cite{14}:
\begin{equation}
  \label{3}
  C_1 =-0.324 ~~~~~~~C_2= 1.150 .
\end{equation}

These coefficients are regularization scheme independent at next-to-leading
logarithmic precision as obtained by Buras et al.\cite{15} for 
$\Lambda_{\overline {MS}}^{(4)}=350 $MeV. With the factorization 
hypothesis and after Fierz reordering we obtain from (\ref{1}) and (\ref{2})
\begin{equation}
  \label{4}
  <\pi^+ \pi ^- | {\cal H}_{eff} | \bar B^0 >=  
\frac{4G_F}{\sqrt{2}} V_{ub}V_{ud}^* \left( \frac{1}{N} C_1 +C_2 \right)
<\pi^+ |\bar u  \gamma_\mu L b| \bar B^0> 
<\pi^-|\bar d \gamma ^\mu L u|0>,
\end{equation}
where N is the number of colours.

The current matrix elements in (\ref{4}) are evaluated in terms of form 
factors $F_1$ and $F_0$ for the decay $\bar B^0 \to \pi^+l^- \bar 
\nu _l$ and the 
pion decay constant $f_\pi =0.132 $GeV as follows:
\begin{equation}
  \label{5}
  <\pi^+ |\bar u  \gamma_\mu L b| \bar B^0> =\frac{1}{2}\left 
[(p_B +p_\pi)_\mu - \frac{m_B^2-m_\pi^2}{q^2}q_\mu \right ] 
F_1^{B\to \pi} (q^2) 
+\frac{1}{2}\frac{m_B^2-m_\pi^2}{q^2}q_\mu F_0^{B\to \pi} (q^2) 
\end{equation}
\begin{equation}
  \label{6}
  <\pi^-|\bar d \gamma ^\mu L u|0>=\frac{i}{2}f_\pi q^\mu
\end{equation}
where $q=p_B - p_\pi$. With these definitions we get the following result 
for (\ref{4}):
\begin{equation}
  \label{7}
<\pi^+ \pi ^- | {\cal H}_{eff} | \bar B^0 >=  i
\frac{G_F}{\sqrt{2}} V_{ub}V_{ud}^* \left( \frac{1}{N} C_1 +C_2 \right)
f_\pi (m_B^2-m_\pi^2) F_0^{B\to \pi} (m_\pi^2)
\end{equation}

In (\ref{5}) $F_1 (F_0)$ are the form factors of $1^- (0^+)$ (transverse
(longitudinal)) projections of the vector current. 
At $q^2=0$, we have  $F_0(0) =F_1(0)$ to cancel the pole at $q^2=0$ in 
(\ref{5}). 
But $F_1$ and 
$F_0$ differ for arbitrary $q^2$. In (\ref{7}) we need $F_0^{B\to \pi}$
only for $q^2=m_\pi^2$, which is small, so that we can set $F_0^{B\to \pi} =
F_1^{B\to \pi}$. 
Of course for $F_0^{B\to \pi}(q^2) = 
F_1^{B\to \pi} (q^2)$ with arbitrary 
$q^2$ in (\ref{5})
we obtain the same result. 
The $1^-$ ($0^+$) part of the current ($\bar u \gamma_\mu L b$) has the 
quantum number of the $B^{*-}$ ($B_0^-$) meson with mass $m_{B^*}=5.32$GeV
($m_{B_0}=5.73$GeV) (We denote the $^3P_J$ bound states by $B_J$).
If according to the vector-meson-dominance model (VDM) the form factor
$F_1^{B\to \pi}$ is dominated by just this $1^-$ resonance with mass $m_{B^*}$
we have
\begin{equation}
  \label{8}
  F_1^{B\to \pi} (q^2) = \frac{g_{B^*B\pi} m_{B^*} f_{B^*}}
{m_{B^*}^2 -q^2}.
\end{equation}
In (\ref{8}) the current coupling $f_{B^*}$ is defined by the vector 
current matrix element
\begin{equation}
  \label{9}
<0| \bar u \gamma_\mu b | \bar B^* >=  f_{B^*} m_{B^*} \epsilon _\mu,
\end{equation}
and  the strong coupling is defined through the Lagrangian 
\begin{equation}
  \label{10}
{\cal L}_ {B^*B\pi} =  g_{B^*B\pi} B^*_\mu B \stackrel {\leftrightarrow}{
 \partial^\mu} \pi .
\end{equation}
Similarly we define the current matrix element of the scalar resonance 
state $B_0^-$:
\begin{equation}
  \label{11}
<0| \bar u \gamma^\mu b | B_0^- >= i f_{B_0} p_{B_0} ^\mu,
\end{equation}
and the coupling of $B_0^-$ to $\bar B^0 \pi^+$ by
\begin{equation}
  \label{12}
{\cal L}_ {B_0 B\pi} = \frac{1}{m_{B_0}} g_{B_0 B\pi} \partial _\mu B_0
 B \stackrel {\leftrightarrow}{
 \partial^\mu} \pi ,
\end{equation}
$g_{B_0 B\pi}$ is dimensionless.
With these definitions the form factor $F_0^{B\to \pi} (q^2)$ is given 
in the scalar-meson-dominance model by the formula
\begin{equation}
  \label{13}
  F_0^{B\to \pi} (q^2) = \frac{g_{B_0B\pi} m_{B_0} f_{B_0}}
{m_{B_0}^2 -q^2}.
\end{equation}
With (\ref{13}) the weak transition matrix element (\ref{7}) can be written
as 
\begin{equation}
  \label{14}
<\pi^+ \pi ^- | {\cal H}_{eff} | \bar B^0 >=  i
\frac{G_F}{\sqrt{2}} V_{ub}V_{ud}^* \left( \frac{1}{N} C_1 +C_2 \right)
f_\pi \frac{m_B^2-m_\pi^2}{m_{B_0}^2-m_\pi^2} g_{B_0B\pi} m_{B_0} f_{B_0}.
\end{equation}
Here $g_{B_0B\pi}$ and $f_{B_0}$ are on-shell couplings to the scalar resonance
 $B_0$ which is on the mass shell. 
This means that the off-shell extrapolation 
factor of the current matrix element is given by just a single pole 
in (\ref{13}).

\begin{figure}
\epsfig{file=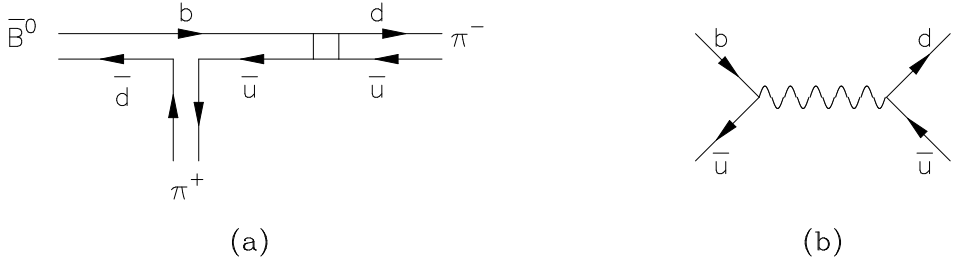,bbllx=4cm,bblly=16cm,bburx=22cm,bbury=21cm,%
width=15cm,angle=0}\label{ff2}
\caption{Feynman diagrams for $\bar B^0 \to \pi^+\pi^-$ in pole model.}
\end{figure}

The result (\ref{14}) for the transition amplitude  $\bar B ^0 \to \pi^+
\pi^-$ can be obtained more directly from the pole diagram in Fig.1a.
In this diagram, the initial $\bar B^0$ can go into a $B^{*-}\pi^+$ or 
 $B_0^{-}\pi^+$ via the strong interaction couplings (\ref{10}) or 
(\ref{12}). The $B^{*-}$ or $B_{0}$
converts into a $\pi^-$ via the weak interaction (\ref{1})
shown in terms of quark lines in Fig.1b. 
The weak matrix element is evaluated in the vacuum insertion approximation 
with the following result for the intermediate $B_0^-$:
 \begin{eqnarray}
<\pi ^- | {\cal H}_{eff} | B_0^{-} > &=&  
\frac{4 G_F}{\sqrt{2}} V_{ub}V_{ud}^* \left( \frac{1}{N} C_1 +C_2 \right)
<\pi^- | \bar d \gamma^\mu L u| 0> <0| \bar u \gamma_\mu Lb|B_0^{-}>
\nonumber\\
& =&- \frac{ G_F}{\sqrt{2}} V_{ub}V_{ud}^* \left( \frac{1}{N} C_1 +C_2 \right)
f_\pi   m_{B_0}^2 f_{B_0}. \label{14p}
\end{eqnarray}
In (\ref{14p}) $p_{\pi^-} =p_{B_0^-}$ and $p_{B_0^-}^2=m_{B_0^-
 }^2$
and the current matrix elements in (\ref{6}) and (\ref{11}) have been used. 
Inserting the propagator of the intermediate $B_0^{-}$ meson
and the strong vertex 
the result for the decay matrix element is:
 \begin{equation}
  \label{15}
<\pi ^+ \pi ^- | {\cal H}_{eff} | \bar B^{0} >= i 
\frac{G_F}{\sqrt{2}} V_{ub}V_{ud}^* \left( \frac{1}{N} C_1 +C_2 \right)
f_\pi \frac{m_B^2 -m_\pi^2}{m_{B_0}^2-m_\pi^2} g_{B_0B\pi}  m_{B_0} f_{B_0}.
\end{equation}
The result 
(\ref{15}) agrees with (\ref{14}) as to be expected. 
To achieve this, two ingredients were essential.
First, the coupling $B_0\to B\pi$ in (\ref{12}) has such a form that 
it vanishes for $m_B^2 =m_\pi^2$, i.e. the coupling of a scalar meson 
to two pseudoscalar mesons of equal masses would be different. 
Second, in (\ref{14p}) we have set $ p_{\pi^-} \cdot p_{B_0^-} = 
p_{B_0^-}^2=m_{B_0^- }^2$. 
This means that the weak transition matrix element $B_0^- \to \pi^-$ is 
evaluated at the on-shell point of the intermediate $B_0^-$ resonance, 
whereas the propagator is calculated at $(p_{\bar B^0} -p_{\pi^+})^2=
p_{\pi^-}^2 =m_\pi^2$.
This is in the spirit of the meson-dominance approximation and agrees 
with the result (\ref{13}) where the form factor $F_0^{B\to \pi} (q^2)$ 
is written down directly in this approximation. 
 
In the pole model calculations of ref.\cite{11} a different route 
has been taken. 
Instead of the $B_0$ state the intermediate state is the $B^*$ with spin $1^-$.
In this case the pole cancels by the factor $m_{B^* }^2 -q^2$ in the 
numerator and the result is 
 \begin{equation}
  \label{16}
<\pi ^+ \pi ^- | {\cal H}_{eff} | \bar B^{0} >=  i
\frac{G_F}{\sqrt{2}} V_{ub}V_{ud}^* \left( \frac{1}{N} C_1 +C_2 \right)
f_\pi \frac{(m_{B}^2 -m_\pi^2)}{m_{B^*}} g_{B^*B\pi} f_{B^*}.
\end{equation}
The difference between (\ref{15}) and (\ref{16}) is small since 
 $F_1^{B\to \pi}(0) =F_0^{B\to \pi} (0) \simeq F_0^{B\to \pi} (m_\pi ^2) $. 
It is clear however, that in this evaluation the residue of the pole
$1/(m_{B^*}^2 -q^2)$ vanishes like $(m_{B^*}^2-q^2)$, so that the pole
is canceled. This is apparent also from (\ref{14p}) when we evaluate 
$<\pi^- | {\cal H}_{eff} |B^{*-}> $ in the vacuum saturation approximation.
Then we obtain the factor $\epsilon \cdot p_{\pi^-} =\epsilon \cdot p_{B^*}$
where  $\epsilon$ is the polarization vector of the $B^*$. 
Of course, this factor vanishes on the mass shell. 
We think that our evaluation of the pole model is more appropriate 
since the pole is kept, so that our result is correct for $q^2 \to m_{B_0}^2$.
Since a large extrapolation from $q^2=m_{B_0}^2$ to $q^2=m_\pi^2$ is involved,
one must be careful in the calculation of the numerator.

  From (\ref{13}) we have  
  \begin{equation}
    \label{17}
    g_{B_0 B\pi} = \frac{F_0^{B\to \pi}(0) m_{B_0}}{f_{B_0}}.
  \end{equation}
and a similar equation for  $g_{B^* B\pi}$ following from (\ref{8}).

 $F_1^{B\to \pi}(0) =F_0^{B\to \pi} (0)$ has been calculated in various 
approaches: quark model, QCD sum rule and QCD lattice calculations. 
Thus either from these results or in case the branching ratio for the decay 
$\bar B^0 \to\pi^+ \pi^-$ becomes precisely known experimentally the relation
(\ref{15}) allows us to calculate the coupling constant $g_{B_0 B\pi}$, 
if $f_{B_0}$ is determined from other sources. 
Similarly $g_{B^* B\pi}$ is calculated from $F_1^{B\to \pi} (0) $, where 
we need in addition the current coupling $f_{B^*}$. 
Since the experimental branching ratio of $\bar B^0 \to\pi^+ \pi^-$ 
is not very well known yet (see Table~5) we must rely on 
$F_1^{B\to \pi}(0) =F_0^{B\to \pi} (0)$ as calculated in various models. 
We shall do this later and compare to direct theoretical results for 
$g_{B_0 B\pi}$ or $g_{B^* B\pi}$.

Our result for the  weak transition matrix elements has the factor 
$a_2 =C_2 +C_1/N$, which should be evaluated with N=3. 
In the following we shall evaluate them with N=2 giving 
$a_2= 0.988$ using (\ref{3}). 
$a_2$ depends on N only mildly, since $|C_1|< C_2$, so that choosing 
N=3 would give similar results for all colour unsuppressed decays, for 
which the transition matrix elements are proportional to $a_2$.
In the following we shall denote this class of decays by class I. 
The class II stands for the colour suppressed decays which have 
matrix elements proportional to $a_1 =C_1 +C_2/N$, which depends strongly on N.
The third class is a superposition of class I and class II matrix elements. 
Recent comparisons of experimental data for many two-body B decays show
that on the basis of the factorization hypothesis and BSW matrix elements from
\cite{13} a reasonable fit to these data gives $N\simeq 2$ \cite{16}. 
N=3 would produce very bad results for the measured class II decays
and the class III decays can be fitted only with a positive $a_1$. 
For example, for N=2, 3 or $\infty$ we have the following values for $a_1$
using (\ref{3}):
$a_1=0.251,0.059,-0.324$.
The choice $N=\infty$ is favored in D decays \cite{7}. The fact, that the 
choice N=3 is not possible points into the direction that so-called 
non-factorization contributions are significant, in particular in the 
class II and class III transitions. Of course, it is not certain, that 
these contributions can universally accounted for by choosing N=2. It 
is more likely that these contributions depend on the particular decay channel.
 For fits and constraints on the non-factorization terms in various 
channels see \cite{17}. 
We also emphasize that the result $N\simeq 2$ from the fit in \cite{16}
depends on the BSW \cite{13} choice of transition form factors and current 
coupling constants. 
We shall assume $N=2$ independent in which pole model the weak matrix elements
are evaluated. 
This can be justified empirically since the magnitude and also the sign of 
$a_1$ is deduced from such final states, where the two pole models coincide,
as will be seen later in our presentation for the PV and VV results. 
If one gives up the requirement that the $C_i$ should be independent of 
the regularization scheme one can find short distance coefficients $C_1$
and $C_2$ in next-to-leading logarithmic precision which give $a_1=0.2$
with N=3 by selecting the appropriate scheme \cite{18}.

\begin{figure}[htb]\label{f3}
\epsfig{file=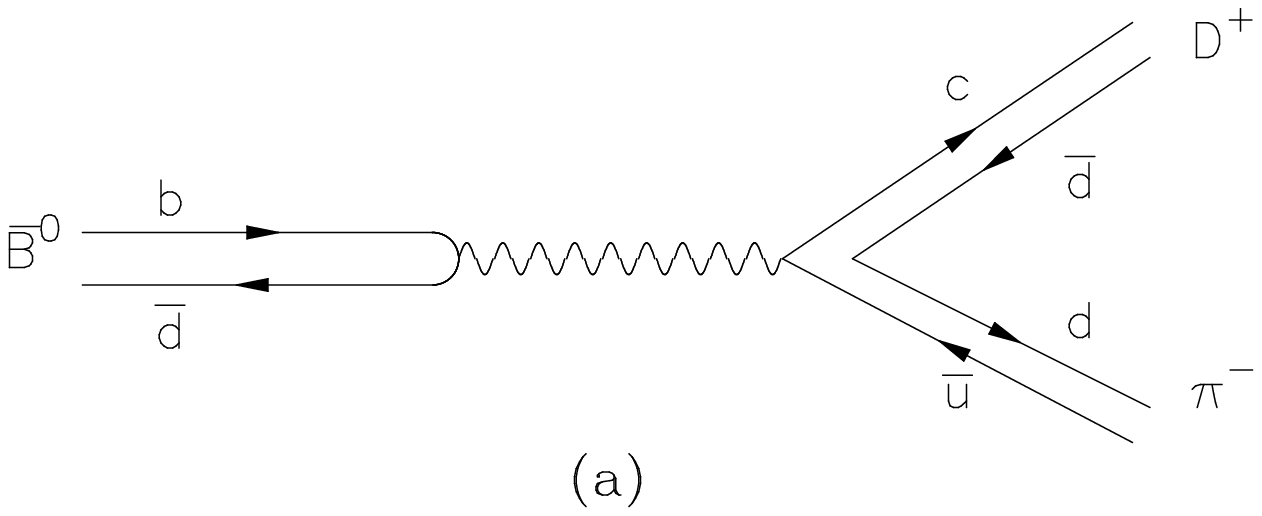,bbllx=4cm,bblly=13cm,bburx=22cm,bbury=19cm,%
width=11.5cm,angle=0}
\epsfig{file=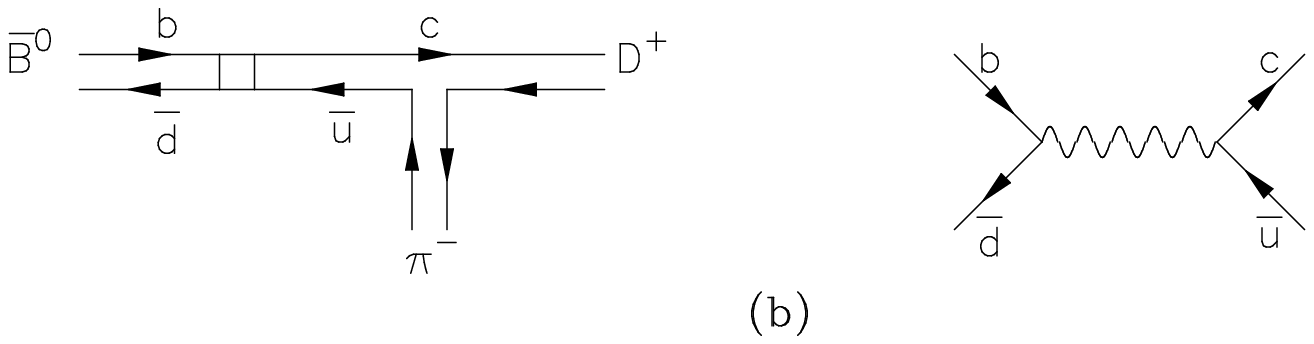,bbllx=4cm,bblly=15cm,bburx=22cm,bbury=19cm,%
width=15.5cm,angle=0}
\caption{The diagrams of $\bar B^0 \to D^+\pi^- $ in the usual 
factorization approximation for 
the quark diagram (a) and for the pole model (b).}
\end{figure}

Other class I decays going into two pseudoscalar mesons are $\bar B^0
\to D^-D^+$, $D_s^-D^+$, $K^-D^+$, $\pi^-D^+$ and
 $ B^- \to D^-D^0$, $D_s^-D^0$. 
These decays are calculated from ${\cal H}_{eff}$'s where $u\to c$ 
and/or $d\to s$ with the same short distance coefficients as in (3).
Of special interest is the decay $\bar B^0 \to \pi ^- D^+$ which belongs
also to the colour unsuppressed class I. This decay can proceed also through 
the annihilation diagram. The diagrams in the usual factorization 
approximation for the quark diagram and for the pole model are shown in 
Fig.\ref{f3}a and \ref{f3}b.

In the pole model according to ref.\cite{11}, 
the transition matrix element corresponding to the 
diagram in Fig.\ref{f3}b is 
 \begin{equation}
  \label{16p}
<\pi ^- D^+| {\cal H}_{eff} | \bar B^{0} >=  -i
\frac{G_F}{\sqrt{2}} V_{cb}V_{ud}^* \left( \frac{1}{N} C_2 +C_1 \right)
f_{B^0} \frac{(m_{D}^2 -m_\pi^2)}{m_{D^*}^2} g_{D^{*0}D^+ \pi^-} 
f_{D^*} m_{D^*} .
\end{equation}
With (\ref{8}) this can be written as 
 \begin{equation}
  \label{17p}
<\pi ^- D^+| {\cal H}_{eff} | \bar B^{0} >=  -i
\frac{G_F}{\sqrt{2}} V_{cb}V_{ud}^* a_1
f_{B^0} (m_{D}^2 -m_\pi^2)  F_1^{D\to \pi} (0).
\end{equation}
Compared to the dominant pole contribution, which is given by the 
analogous equation to (\ref{14}), the annihilation contribution is 
suppressed by the factor $\frac{m_D^2}{m_B^2}\frac{a_1}{a_2}$ which is 
approximately 0.03.

The fact that in (\ref{17p}) the pole $1/(q^2 -m_{D^*}^2)$ has canceled 
against the same factor $(q^2-m_{D^*}^2)$ in the numerator indicates that 
the residue of the pole vanishes on the mass shell of the $D^{*0}$ resonance. 
This is already obvious when one evaluates the weak transition matrix 
element in the vacuum saturation approximation
 \begin{equation}
  \label{18p}
<D^{*0} | {\cal H}_{eff} | \bar B^{0} > =  
\frac{4G_F}{\sqrt{2}} V_{cb}V_{ud}^* a_1 <D^{*0} | \bar c \gamma_\mu L u 
 | 0 > <0 | \bar d \gamma^\mu L b   | \bar B^{0} >.
\end{equation}
The first current matrix element in (\ref{18p}) is proportional to the 
polarization vector of the $D^{*0}$ whereas the second current matrix element
is proportional to $p_{B^0}^\mu = p_{D^{*0}}^\mu$, so that the product 
vanishes.
This indicates, that the pole-model calculation of the annihilation 
contribution is also ambiguous.

If we start from the quark diagram in Fig.\ref{f3}a and do the vacuum 
insertion 
approximation, but then evaluate the matrix element $ <\pi ^- D^{+} 
| \bar c \gamma_\mu L u  | 0 >$ directly in terms of the form factor $ 
F_0^{D\to \pi} (q^2)$ we obtain the usual result:
 \begin{eqnarray}
  \label{18}
<\pi ^- D^+| {\cal H}_{eff} | \bar B^{0} > &=& - i
\frac{G_F}{\sqrt{2}} V_{cb}V_{ud}^* a_1
f_{B^0} (m_{D}^2 -m_\pi^2)  F_0^{D\to \pi} (m_B^2)\\\nonumber
&=&  -i \frac{G_F}{\sqrt{2}} V_{cb}V_{ud}^* a_1
f_{B^0} (m_{D}^2 -m_\pi^2)  F_0^{D\to \pi} (0) \frac{ m_{D_0}^2}{m_{D_0}^2
-m_B^2}.
\end{eqnarray}
This means that the annihilation contribution is suppressed further by the
additional factor $m_{D_0}^2/(m_{B}^2 -m_{D_0}^2)$, which results from 
the fact 
that in (\ref{18}) the current transition form factor is highly off-shell at 
$q^2=m_B^2$, whereas the off-shell extrapolation in the pole diagram is 
determined by the mass of the $D^{*0}$.
It is clear that the pole-model calculation starting from Fig.\ref{f3}a 
with a  $0^+$ intermediate state leads to the same result (\ref{18}).

\begin{table}\begin{center}
\caption{Values of pole masses in GeV.}
\begin{tabular} {|c|c|c|c|c|}
\hline
Current & m($0^-$) & m($1^-$) &  m($1^+$)&  m($0^+$) \\
\hline
$\bar d c$ & 1.8693 & 2.010 & 2.4222& \\
$\bar u c$ & 1.8645 & 2.0067 & 2.4222&\\
$\bar s c$ &  1.9685 & 2.1124 & 2.53535&\\
$\bar u b$ & 5.2789 & 5.3248 &5.37& 5.73\\
$\bar d b$ & 5.2792 & 5.3248 &5.37& 5.73\\
$\bar s b$ & 5.3693 & 5.41 &5.9 & 5.89\\
$\bar c b$ & 6.264  & 6.337 & 6.730 & 6.700\\
$\bar c c$ &2.9798 & 3.09688 &3.51053& 3.4151\\
\hline
\end{tabular}\end{center}
\end{table}

Such an annihilation contribution is in principle also possible for 
$\bar B^0 \to \pi^+ \pi^-$. But it vanishes since the particles in the 
final state have equal masses.

We conclude that the pole-model calculation of the non-annihilation diagrams
 as depicted in Fig.1a agrees with the usual calculations of the
$B\to PP$ decays where the form factors are approximated  by a single pole
(compare (\ref{14}) with (\ref{15}) ). If a different off-shell extrapolation 
as advocated in ref.\cite{11} is assumed we obtain the result (\ref{16}) which 
differs actually very little from (\ref{15}) 
 since  $F_0^{B\to \pi} (0)=F_1^{B\to \pi} (0) $ and 
the extrapolation of $F_0^{B\to \pi} (q^2)$ from $q^2 =0$ to $q^2=m_\pi^2$
 has only a small effect.
Of course, this is different for decays like $\bar B^0 \to D^-D^+, D_s^-D^+$,
 for which the form factor $F_0^{B\to D} (q^2)$ must be computed for 
$q^2=m_D^2$, which
differs from $F_1^{B\to D} (0)$ to be inserted into (\ref{16}). 

Class II decays are $\bar B^0 \to \pi^0  \pi^0,  \pi^0  D^0$, $\bar
K^0 \eta_c$, and 
$B^-\to K^- \eta_c$. The decays  $B^-\to \pi^-  \pi^0, \pi^-  D^0$ and 
$K^-  D^0$ are class III decays, where contributions of colour 
suppressed and unsuppressed matrix elements interfere.

To obtain an overview on expected branching ratios for all these
decays we shall present results in two schemes:
(i) the pole-model calculation in the form  (\ref{14}) which is equal to the 
usual quark diagram computation in terms of form factors $F_0 (0)$; 
(ii) the pole model result (\ref{16}) where the pole canceled and the 
result depends on $F_1(0)$.
In scheme (i) the annihilation contribution which occurs in some channels 
is negligible, either due to strong suppression in the quark diagram result 
(\ref{18}) or due to the same suppression for the pole model. 
In scheme (ii) we evaluate the annihilation diagram with the formula 
(\ref{17p})
which involves a different off-shell extrapolation of the residue of the pole. 
It is clear that in this form the annihilation terms are larger. The scheme 
(i) will be denoted BSW model since the pole model calculation agrees with 
the usual computation and the scheme (ii) will be denoted pole model. 

\begin{table}\begin{center}
\caption{Values of decay constants. }
\begin{tabular} {|cc|cc|}
\hline
 Particle & $f_M$ (MeV) & Particle & $f_V$ (MeV) \\
\hline
$\pi$ & 133 & $\rho$ & 221\\
 K  & 162 & $K^*$  & 221\\
D   & 250 & $D^*$  & 250\\
$D_s$  & 288 & $D_s^*$  & 288 \\
 B  & 140 & $B^*$  & 160\\
 $B_s$  & 158 & $B_s^*$  & 181 \\
 $B_c$  & 360 & $B_c^*$  & 320 \\
$\eta_c$  & 393 & $J/\psi$ & 382\\
\hline
\end{tabular}\end{center}
\end{table}

For these calculations we need the values of the  masses of the particles 
in the initial and final states and of the pole positions together with
 the pseudoscalar ($f_M$) and vector meson ($f_V$) decay constants. 
The masses are taken from the PDG  tables \cite{19} or from the table
in \cite{7}. 
The decay constants $f_M$ and $f_V$ for pseudoscalar and vector mesons, 
respectively are taken from the following sources. 
$f_\pi$, $f_K$, $f_\rho$ and $f_{K^*}$ are taken from \cite{7}. 
$f_{D_s} = 288$MeV corresponds to the CLEO measurement \cite{Z} of the 
leptonic decay of the $D_s^-$.
We assume $f_{D_s^*}=f_{D_s}$ and fix $f_D$ by the ratio $f_{D_s}/ f_D =1.15$,
which is the average value of this ratio obtained in various lattice 
calculations \cite{s16}. 
$f_D = f_{D^*}$ is also fixed by assumption. 
$f_B$ and $f_{B^*}$ are from QCD sum rule calculations in \cite{25}. 
The values for $f_{B_s}$ and $f_{B_s^*}$ are fixed by the ratio 
 $f_{B_s}/f_{B}=f_{B_s^*}/f_{B^*}=1.13$ from \cite{s16}. The values for 
$f_{B_c}$ and $f_{B_c^*}$ have been found in \cite{27}. The ratio of 
$f_{\eta_c}^2/f_{J/\psi}^2=1.06$ is found in
\cite{kim} and $f_{J/\psi}$ is determined from the leptonic width of the 
$J/\psi$ as in \cite{7}.
Of course, most of these values for the decay constants are not definite yet
 and somewhat other choices can be made.
This information is collected in Table~1 (masses) and Table~2 (decay 
constants). 
The form factors $F_1(0)=F_0(0)$ are written down in Table~3, also taken 
from \cite{7}. 
The CKM matrix elements are given in Table~4, together with the $B^0$ 
and $B^-$ lifetime data \cite{19}.
Our results for the branching ratios in the pole model and the BSW model 
are collected in Table~5. 
In this table we also give the amplitudes for the BSW and 
 for the pole model which show 
in particular the contributions proportional to $a_1$ and $a_2$
for the class III decays. 
For some of the channels the branching ratios have been measured. The 
world-average values are reported in \cite{16}. 
For other channels some new measurements and 
upper limits exist \cite{20}. 
These data are given in the last column of Table~5.

\begin{table}\begin{center}
\caption{Form factors at zero momentum transfer \cite{7}. }
\begin{tabular} {|c|ccccc|}
\hline
Decay & $F_1 =F_0$ & V & $A_1$ & $A_2$ & $A_0$\\
\hline
$D \to \pi$ & 0.692 &      &       & &\\
$D \to K$ & 0.762 &      &       & &\\
$D \to \rho$ &      & 1.225 & 0.775 &  0.923 & 0.669\\
$D^* \to \pi$ &      & 1.225 & 0.775 &  0.923 & 0.669\\
$D_s \to K  $ & 0.643 &      &       & &\\
$B \to \pi$ & 0.333 &      &       & &\\
$B \to K$   & 0.379 &      &       & &\\
$B \to D$   & 0.690 &      &       & &\\
$B \to \rho$ &      & 0.329 & 0.283 &  0.283& 0.281\\
$B \to K^*$ &      & 0.369 & 0.328 &  0.331 & 0.321\\
$B \to D^*$ &      & 0.705 & 0.651 &  0.686& 0.623\\
\hline
\end{tabular}\end{center}
\end{table}

\begin{table}
  \begin{center}
  \caption{CKM matrix elements and life time of B mesons.}
    \leavevmode
    \begin{tabular}{|c|cccccc|c|c|}
\hline
 Name &  $|V_{ud}|$ &  $|V_{cs}|$ & $|V_{us}|$ &$|V_{cd}|$ & $|V_{cb}|$
 &$|V_{ub}|$ & $\tau_{B^0}$ (s)& $\tau_{B^-}$ (s)\\
\hline
Values& 0.975&0.9742 & 0.221&0.221& 0.038&0.0035&$1.56\times 10^{-12}$& 
$1.62\times 10^{-12}$\\
   \hline
 \end{tabular}
  \end{center}
  \label{tab:4}
\end{table}

\begin{small}
\begin{table}\begin{center}
\caption{$B\to PP$ Amplitudes($10^{-7}$GeV) and Branching Ratios 
in two versions of pole models and experimental measured branching ratios
or upper limits (90\%C.L.) \cite{16,20}. }
\begin{tabular} {|l|cc|cc|c|}
\hline
 & \multicolumn{2}{|c|}{Pole model} & \multicolumn{2}{|c|}{BSW model} &
Experimental \\
Channel & Amplitude &  BR &Amplitude &  BR &  BR \\
\hline
 $B^- \to D^-  D^0$  & 2.91 $a_2$ & $5.44\times 10^{-4}$ & 
 3.16 $a_2$ &$6.39\times 10^{-4}$&-\\
$B^- \to D_s^-  D^0$  & 14.8   $a_2$ & $1.36\times 10^{-2}$ &
 16.2   $a_2$ & $1.63\times 10^{-2}$ &($1.36\pm 0.43)
10^{-2}$\\
$\bar B^0 \to \pi^-  \pi^+$  & 0.35  $a_2$ & $1.05\times 10^{-5}$  & 
 0.35  $a_2$ & $1.05\times 10^{-5}$ & $ (0.7\pm 0.4)
10^{-5}$\\
$\bar B^0 \to K^- D^+$  & 1.89 $a_2$ & $2.68\times 10^{-4}$ & 
 1.90 $a_2$ &$2.71\times 10^{-4}$ &-\\
$\bar B^0 \to \pi^- D^+$  & 6.84$a_2$-1.03$a_1$ & $3.29\times 10^{-3}$ &
 6.84 $a_2$ &$3.56\times 10^{-3}$ &(
$3.1\pm 0.4) 10^{-3}$\\
$\bar B^0 \to D_s^- D^+$  & 14.8   $a_2$ &$1.31\times 10^{-2}$  &
 16.2   $a_2$ &$1.57\times 10^{-2}$ &$(0.74\pm 0.28) 10^{-2}$\\
$\bar B^0 \to D^- D^+$  & 2.91   $a_2$ &  $5.22\times 10^{-4}$ & 
 3.16   $a_2$ &   $6.14\times 10^{-4}$& $< 1.2\times 10^{-3}$
\\
\hline
$ B^- \to K^- \eta_c$  & 12.6   $a_1$ & $6.13\times 10^{-4}$ &
 16.9   $a_1$ & $1.11\times 10^{-3}$ &-\\
$\bar B^0 \to \bar K^0 \eta_c$  & 12.6   $a_1$ &  $5.90\times 10^{-4}$ &
 16.9   $a_1$ &  $1.07\times 10^{-3}$  &-
\\
$\bar B^0 \to \pi^0 \pi^0$  & 0.17   $a_1$ &$3.39\times 10^{-7}$ &
 0.17   $a_1$ &$3.39\times 10^{-7}$ &$<9.1\times 10^{-6}$
\\
$\bar B^0 \to \pi^0 D^0$  & 5.01$a_1$-0.72$a_1$ & $9.04\times 10^{-5}$ &
 5.60 $a_1$ &$1.54\times 10^{-4}$ &$<4.8\times 10^{-4}$\\
\hline
$ B^- \to K^- D^0$  & $1.89a_2$+$1.81a_1$ &  $4.32\times 10^{-4}$ &
$1.90a_2$+$2.02a_1$ & $4.55\times 10^{-4}$ &-
\\
$ B^- \to \pi^- D^0$  & $6.84 a_2$+7.08$a_1$ &$5.91\times 10^{-3}$  &
 $6.84 a_2$+7.92$a_1$ & $6.20\times 10^{-3}$ &
$(5.0\pm 0.5)10^{-3}$\\
$ B^- \to \pi^- \pi^0$  & $0.25a_2$+0.25$a_1$ &$8.57\times 10^{-6}$ &
$0.25a_2$+0.25$a_1$ & $8.58\times 10^{-6}$ & $(9^{+6}_{- 5})
 \times 10^{-6}$\\
\hline
\end{tabular}\end{center}
\end{table}
\end{small}

\begin{table}\begin{center}
\caption{The form factor $F_1^{B\to \pi}(0)$ at $q^2=0$ in different models. }
\begin{tabular}{|cc|c|}
\hline
Reference & method & $F_1^{B\to \pi}(0)$ \\
\hline
\cite{7} & a & 0.333 \\
\cite{22} & a & 0.293 \\
\cite{9} & a & $0.21\pm 0.02$ \\
\cite{t5} & b & 0.53 (0.89) \\
\cite{t6}& c & 0.29$\pm 0.01$\\
\cite{t7}& c & $0.26\pm 0.02$ \\
\cite{t8} & c & $0.26\pm 0.02$ \\
\cite{t9} & c & $0.23\pm 0.02$ \\
\cite{t10} & c & $0.4\pm 0.1$ \\
\cite{t11} & c & $0.24$ \\
\cite{21} & d & $ 0.37\pm 0.05$ \\
\cite{t12} & e & 0.10-0.49 \\
\cite{t13} & e & 0.23-0.43 \\
\cite{t14} & e & 0.21-0.27 \\
\hline
\multicolumn{3}{l}{a Quark model}\\
\multicolumn{3}{l}{b HQET and CPTT}\\
\multicolumn{3}{l}{c QCD sum rules}\\
\multicolumn{3}{l}{d experimental value from $B( B^0 \to \pi ^- l^+ \nu)$ }\\
\multicolumn{3}{l}{e Lattice calculations. }
\end{tabular}\end{center}
\end{table}

 From the results in Table~5, we observe the following. 
Pole model and BSW results lie very near together. Exceptions are the decays
$B^- \to D^- D^0$, $D_s^-D^0$ and $\bar B^0 \to D^-D^+$, $D_s^-D^+$ 
which differ due to the extrapolation of $F_0^{B\to D}$ from $q^2=0$
to $q^2=m_D^2$. 
The annihilation contribution in $\bar B^0 \to \pi^- D^+$ has little effect.
However, it is significant for the decay $\bar B^0 \to \pi^0 D^0$, where it 
is proportional to the same Wilson coefficient $a_1$ as the 
direct contribution. Except for the decay $\bar B^0\to D_s^- D^+$ we observe
very good agreement between the calculated and the measured branching ratios. 
Both pole model versions agree with the data, so that they can not be
 utilized to exclude for example the model based on ref.\cite{11}.

For the results in Table~5 we needed the values of three form factors 
$F_1^{B\to D}$, $F_1^{B\to \pi}$ and $F_1^{B\to K}$ taken from Table~3. 
Since only for four channels, namely $B^-\to D_s^-D^0$, $\pi^-D^0$ 
and $\bar B^0 \to D_s^-D^+$, $\pi^-D^+$, experimental branching ratios exist,
we gain only information on $F_1^{B\to D}(0)$ and  $F_1^{B\to \pi}(0)$.
Since our results agree quite well with the measured data we are 
confident that the BSW values for these form factors are reasonable. 
In this connection it is of interest to search in the literature for 
other theoretical predictions on these form factors. 
Such predictions based on various methods are collected in Table~6 for 
 $F_1^{B\to \pi}(0)$ and in Table~7 for $F_1^{B\to D}(0)$.
The values for  $F_1^{B\to \pi}$ with two exceptions are very similar. 
Calculations using similar methods give approximately the same result. 
Furthermore the values predicted from QCD sum rules and constituent quark
models of various kinds are very near together and agree with the 
average value obtained from QCD lattice calculations. 
 From the comparison in Table~6 we are confident that $F_1^{B\to \pi}(0)
\simeq 0.3$ is a reasonable value which agrees also with the recent  
measurement of the branching ratio for $B^0\to \pi^- l^+\nu$ \cite{21}. 
$F_1^{B\to \pi}(0)=0.37\pm 0.05$ is deduced from the measured branching 
ratio, using $|V_{ub}|=0.0035$ from Table~4 and the pole masses from Table~1. 
This value depends on the way the $q^2$ dependence of the form factor 
is parameterized. We assumed for consistency the approximation with one pole.
Results with other form factor assumptions can be found, for example, in 
\cite{22}.

\begin{table}\begin{center}
\caption{The form factor $F_1^{B\to D}(0)$ at $q^2=0$ in different models. }
\begin{tabular} {|cc|c|}
\hline
Reference & method & $F_1^{B\to D}(0)$ \\
\hline
\cite{7} & a & 0.690 \\
\cite{22} & a & 0.684 \\
\cite{t15} & a & 0.63 \\
\cite{t9} & c & $0.62\pm 0.06$ \\
\cite{23} & f & 0.74 \\
\cite{19} & g & $0.71\pm 0.10$ \\
\hline
\multicolumn{3}{l}{f HQET with monopole form factor and $\xi(1)=0.91$ 
\cite{24}. }\\
\multicolumn{3}{l}{g experimental value from $B( B^0 \to D ^- l^+ \nu)$. }
\end{tabular}\end{center}
\end{table}

The situation is similar for the form factor $F_1^{B\to D} (0)$. 
The results obtained from quark models agree quite well and agree 
nicely with the value calculated from the measured branching ratio $B(B^0
\to D^-l^+\nu _l )= (1.9\pm 0.5 ) \times 10^{-2}$ \cite{19}.
Of course, the experimental value $F_1^{B\to D} (0)= 0.71 \pm 0.10$ and also
the results obtained from the Isgur -Wise normalization \cite{23} in the 
HQET depend on the assumed $q^2$ variation of $F_1^{B\to D}$. 
The results in Table~7 under f and g are obtained with single-pole
approximation with masses from Table~1. 
Somewhat smaller values as from the extrapolation with 
the single-pole formula
are obtained with the usual exponential extrapolation to $q^2=0$ on the 
basis of slope calculations from QCD sum rules or lattice calculations
or from fits to the experimental data of 
$B\to D^* l\nu$ decays \cite{24}. 
Taking into account that the Isgur-Wise normalization is $\xi (1) = 
0.91 \pm 0.04$ \cite{24} due to mass and perturbative QCD effects 
one arrives at $F_1^{B\to D} (0) \simeq 0.61$, instead of $F_1^{B\to D}
 (0) =0.71$ in Table 7. We conclude that the BSW value is a good average.

We are now in the position to calculate $g_{B^*B\pi}$, $g_{B_s^*BK}$ and 
$g_{B_c^*BD}$ from (\ref{17}). 
To be definite we use the BSW values of the corresponding form factors  
from Table~3 together with the decay constants from Table~2. 
We obtain for the strong VPP coupling constants 
$g_{B^*B\pi} =11.1$, $g_{B_s^*BK}=11.7$, and $g_{B_c^*BD}= 13.7$. 
Surprisingly, these three couplings lie close together, 
so that the spread of the form factors $F_1^{B\to P} (0)$ 
($P=\pi$, $K$, $D$) is mostly due to the different values of 
$f_{B^*}$, $f_{B_s^*}$ and  $f_{B_c^*}$ and the different  mass values. 
If we take $f_{B_c^*}=0.400$GeV, i.e. the upper limit of the QCD sum 
rule result \cite{27} we obtain  $g_{B_c^*BD}= 11.0$. 
In potential model calculations $f_{B_c^*}$ comes out even larger: 
$f_{B_c^*} \simeq 0.5$GeV \cite{28}. 
Results for $g_{B^*B\pi}$ based on QCD sum rule calculations 
or on the quark model with chiral HQET are collected in \cite{25}.
Our result $g_{B^*B\pi} =11.1$ can be compared with various QCD sum rule 
calculations. 
The calculations use different approaches. 
They are either based on expansions near the light-cone or use the 
short-distance expansion in connection with the soft pion limit. 
Results obtained with the first method are $g_{B^*B\pi} =14\pm 3 $ \cite{25},
 $g_{B^*B\pi} =10\pm 2 $ \cite{36a},
 $g_{B^*B\pi} =10.4\pm 2.0 $ \cite{36b},
and  $g_{B^*B\pi} =7\pm 2 $ \cite{36c}.
The second method has given the following results:
$g_{B^*B\pi} =14.5\pm 1.5 $ \cite{25} and $g_{B^*B\pi} =11.2\pm 2.0 $ 
\cite{36d}. These results depend on the values obtained for $f_B$ and 
$f_{B^*}$, which are also taken from QCD sum rule calculations. 
In \cite{36a,36b,36d} these constants are $f_B=(150\pm 20) $MeV and 
$f_{B^*} =(190 \pm 10) $MeV. 
This explains, for example, the different results in \cite{25} and \cite{36d}
 with the light-cone sum rule. 
In the original result for $f_B f_{B^*} g_{B^*B\pi}$, as it follows from 
the sum rule, the two evaluations gave the same result, namely
 $f_B f_{B^*} g_{B^*B\pi}=7.9 \times 10^{-3} GeV^2$. 
The smaller value of $g_{B^*B\pi}$, obtained by Narison et al. \cite{36c} 
is due to the fact, that these authors included $O(\alpha_s)$ corrections 
to the heavy meson decay constants which result in larger values for $f_B$ 
and $f_{B^*}$.
Taking all these uncertainties into account the value of $g_{B^*B\pi}$ 
obtained from the non-leptonic decays agrees reasonably well with the 
QCD sum rule results.

To our knowledge there exist no QCD sum rule calculations for 
$g_{B_s^*BK}$ and $g_{B_c^*BD}$ to which we could compare our results. 

Actually in the BSW pole model the decay amplitude depends on the form 
factor $F_0^{B\to \pi} (0) $ (see (\ref{7})) not $F_1^{B\to \pi}(0)$. 
At $q^2=0$ we have $F_0^{B\to \pi}(0) = F_1^{B\to \pi}(0)$ which yields 
an equation for $g_{B_0B\pi} $ in terms of $g_{B^*B\pi} $:
\begin{equation}
  \label{**}
  g_{B_0B\pi} =\frac{m_{B_0}}{m_{B^*}} \frac{f_{B^*}}{f_{B_0}}
g_{B^*B\pi}
\end{equation}
i.e.  $g_{B_0B\pi} \simeq g_{B^*B\pi} $. 
Inserting masses and assuming $f_{B_0} = f_B$, we obtain  $g_{B_0B\pi} =
13.7$. 
This result can be compared with a recent QCD sum rule calculation of 
this coupling constant \cite{36e}.
Adjusting their result to $f_B=f_{B_0} =0.14$GeV and to our definition 
of the $B_0B\pi$ coupling in (\ref{12}) the result in \cite{36e} is  
$g_{B_0B\pi}=4.5$ and  $g_{B_0B\pi}=6.3$ depending on the QCD sum rule 
method used. 
These values are more than a factor of two smaller than our result. 
It is clear that these sum rule results are not compatible with the 
relation (\ref{**}). 
Since our coupling (\ref{12}) and the $B_0B\pi$ coupling used in 
\cite{36e} have different off-shell extrapolations it is also not 
clear to us whether this has any effect on the sum rule evaluation.

We conclude that the coupling constant  $g_{B^*B\pi}$ obtained from 
non-leptonic decay data is in reasonable agreement with QCD sum rule results. 
The coupling constants  $g_{B_s^*B K}$ and  $g_{B_c^*B D}$ have rather 
similar values to  $g_{B^*B\pi}$.
It would be interesting to know whether these relations can be explained 
in the framework of QCD sum rule calculations.

\section{B decays into pseudoscalar and vector and two vector mesons}

\begin{figure}
\epsfig{file=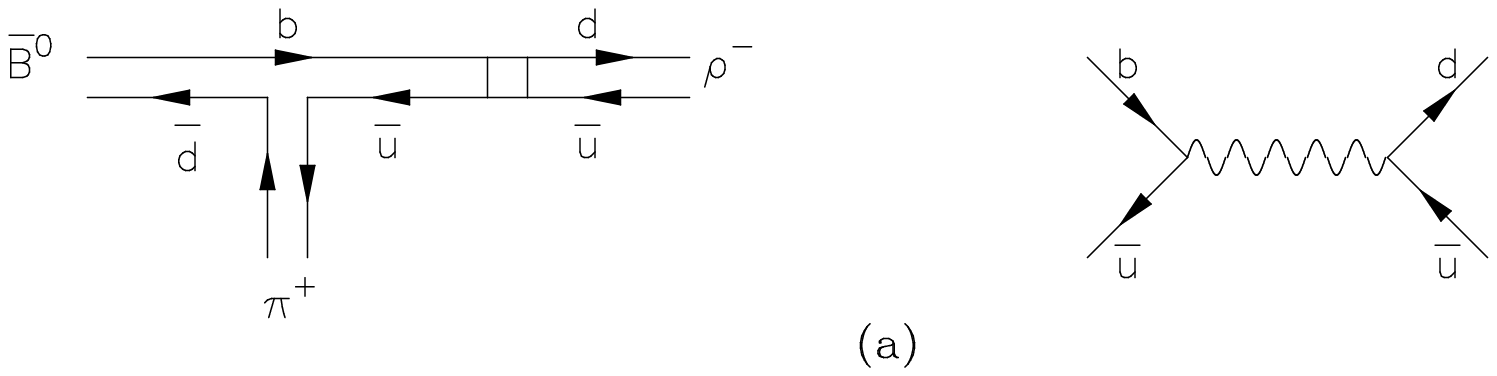,bbllx=4cm,bblly=16cm,bburx=22cm,bbury=21cm,%
width=15cm,angle=0}
\epsfig{file=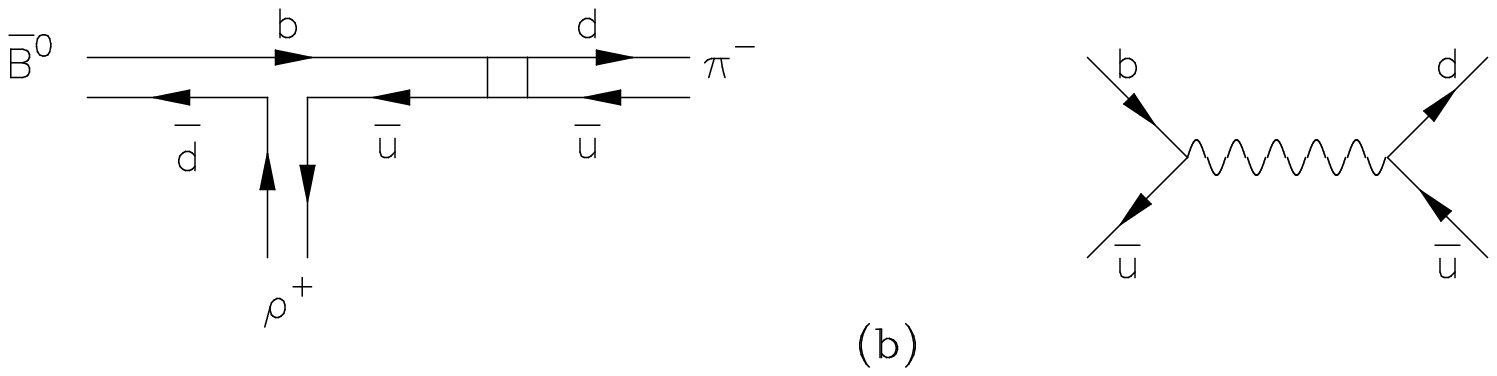,bbllx=4cm,bblly=16cm,bburx=22cm,bbury=21cm,%
width=15cm,angle=0}\label{f4}
\caption{Feynman diagrams for $\bar B^0 \to \pi^+\rho^-$ (a) and 
for $\bar B^0 \to \pi^-\rho^+$ (b) in pole model.}
\end{figure}

First, we consider decays of B mesons into one vector and one pseudoscalar 
meson: $B\to VP$. 
As examples we choose $\bar B^0 \to \pi^+ \rho ^-$ and  $\bar B^0 \to 
\pi^- \rho ^+$, the analogous channels to the decay considered in the 
previous section. The pole model diagram for  $\bar B^0 \to \pi^+ \rho ^-$
is shown in Fig.\ref{f4}a. 
The intermediate state is the $B^*$ resonance which decays weakly into a 
$\rho^-$ meson. 
The result of the pole model evaluation is 
  \begin{equation}
  \label{s1}
<\pi ^+ \rho ^- | {\cal H}_{eff} | \bar B^{0} >=  
\frac{G_F}{\sqrt{2}} V_{ub}V_{ud}^* a_2 m_{\rho} f_{\rho}
\epsilon^*\cdot (p_B+p_\pi)
 \frac{ g_{B^*B\pi} f_{B^*} m_{B^*}} {m_{B^*}^2-m_\rho^2} ,
\end{equation}
which can be written as 
  \begin{equation}
  \label{s2}
<\pi ^+ \rho ^- | {\cal H}_{eff} | \bar B^{0} >=  
\frac{G_F}{\sqrt{2}} V_{ub}V_{ud}^* a_2 m_{\rho} f_{\rho}
\epsilon^*\cdot (p_B+p_\pi) F_1^{B\to\pi}(m_\rho^2) ,
\end{equation}
using the vector-dominance relation (\ref{8}) for the form factor 
$F_1^{B\to\pi}$. This result is identical to the result of the usual 
quark model calculation of BSW. 
It is clear that an intermediate scalar resonance $B_0$ gives a vanishing 
transition matrix element.
So, for this case the two pole model calculations yield a unique result.
The same formula applies to the decays $\bar B^0 \to D^+ D_s ^{*-}$,
$ D^+ D ^{*-}$,  $D^+ \rho ^{-}$ and $D^+ K ^{*-}$.

For the decay $\bar B^0 \to \pi^- \rho ^+$, we need in the usual calculation
from the quark diagram the axial-vector current matrix element between 
$\bar B^0 $ and $\rho^+$. This is parameterized by three invariant 
form factors $A_0$,  $A_1$ and $A_2$ \cite{8}:
  \begin{eqnarray}
<\rho ^+ | \bar u \gamma_\mu \gamma _5 b | \bar B^{0} >&=&  
i\left( \epsilon_\mu^* -\frac{\epsilon^*\cdot q}{q^2}q_\mu\right) 
(m_B+ m_{\rho}) A_1^{B\to\rho} (q^2) \nonumber \\
&-&i\left( (p_B+p_\rho)_\mu -\frac{(m_B^2 -m_{\rho}^2)}{q^2}q_\mu\right)
(\epsilon^* \cdot q) \frac {A_2^{B\to\rho} (q^2) }{m_B+ m_{\rho}}\nonumber \\
&+&i\frac{2m_\rho (\epsilon^* \cdot q)}{ q^2} q_\mu A_0^{B\to\rho} (q^2) ,  
\label{s3}
\end{eqnarray}
where $q=p_B - p_\rho =p_\pi$ and $\epsilon^*$  is the polarization vector 
of the $\rho^+$. 
To cancel the poles at $q^2=0$, we must have 
  \begin{equation}
  \label{s4}2m_\rho  A_0^{B\to\rho} (0)=
(m_B+ m_{\rho}) A_1^{B\to\rho} (0) 
-(m_B-m_{\rho}) A_2^{B\to\rho} (0) .
\end{equation}
With (\ref{s3}) we obtain for the weak transition matrix element the 
usual result:
  \begin{equation}
  \label{s5}
<\pi ^- \rho ^+ | {\cal H}_{eff} | \bar B^{0} >=  
\frac{G_F}{\sqrt{2}} V_{ub}V_{ud}^* a_2  f_{\pi}
(\epsilon^*\cdot p_{\pi^-}) 2 m_{\rho} A_0^{B\to\rho} (m_\pi^2).
\end{equation}
The form factor  $A_0^{B\to\rho} $ can be approximated by a single pole 
with spin $0^-$. Analogously to (\ref{8}) and (\ref{13}) this 
meson-dominance approximation yields 
  \begin{equation}
  \label{s6}
A_0^{B\to\rho} (q^2)=\frac{g_{BB\rho} f_B m_B^2}{(m_B^2-q^2)m_\rho}.
\end{equation}
The intermediate state is the well known $B^-$ particle with mass $m_B$. 
When we evaluate the pole-model diagram in Fig.\ref{f4}b with an intermediate 
 $0^-$ particle directly we obtain the same result, namely
 \begin{equation}
  \label{s7}
<\pi ^- \rho ^+ | {\cal H}_{eff} | \bar B^{0} >=  
\frac{G_F}{\sqrt{2}} V_{ub}V_{ud}^* a_2  f_{\pi}
(\epsilon^*\cdot p_{\pi^-}) \frac{2 g_{BB\rho} f_B m_B^2}{m_B^2-m_\pi^2}.
\end{equation}
For the calculation with a $1^+$ resonance $B_1$ as intermediate state in 
Fig.\ref{f4}b, following ref.\cite{11}, we need the coupling between 
the $\rho$, the $B$ and this resonance. 
This coupling has the following form 
 \begin{equation}
  \label{s8}
{\cal L}_{BB_1\rho} =g_s m_{B_1} BB_1^\mu \rho_\mu 
+\frac{g_d}{m_{B_1}} \left [ B (\partial ^\mu B_1^\nu ) \partial _\nu 
\rho _\mu - (\partial _\nu B)( \partial ^\mu B_1^\nu)\rho _\mu \right], 
\end{equation}
which leads to the following matrix element for the transition $\bar B^0
 \to B_1^- \rho ^+$:
 \begin{equation}
  \label{s9}
<B_1^- \rho ^+ | \bar B^0>  =g_s m_{B_1} \epsilon^*_{B_1}\epsilon^*_\rho
-\frac{g_d}{m_{B_1}} [\epsilon^*_ {B_1}\cdot (p_B+p_\rho)] (\epsilon^*_\rho
\cdot p_{B_1}).
\end{equation}
With these definitions for the two couplings proportional to $g_s$
 (s-wave transition) and $g_d$ (d-wave transition) the weak transition matrix
element becomes 
 \begin{equation}
  \label{s10}
<\pi ^- \rho ^+ | {\cal H}_{eff} | \bar B^{0} >= 
\frac{G_F}{\sqrt{2}} V_{ub}V_{ud}^* a_2  f_{\pi}
(\epsilon^*\cdot p_{\pi^-}) \left[g_s f_{B_1}-
\frac{m_B^2-m_\rho^2}{ m_{B_1}^2}g_d f_{B_1}\right].
\end{equation}
In analogy to (\ref{s6}) we relate the strong couplings in (\ref{s10}) to the 
transition from factors $A_1(0)$ and $A_2(0)$ appearing in the current 
matrix elements of the axial-vector current as defined in (\ref{s3}):
\begin{eqnarray}
g_s f_{B_1} &=& ( m_B + m_\rho)A_1^{B\to \rho}(0) ,\nonumber\\
 g_d f_{B_1} &=& \frac{m_{B_1}^2}{( m_B + m_\rho) }A_2^{B\to \rho}(0) .
 \label{s11}
\end{eqnarray}
Then (\ref{s10}) can be written in terms of $A_1$ and $A_2$ or $A_0$:
\begin{eqnarray}
<\pi ^- \rho ^+ | {\cal H}_{eff} | \bar B^{0} > &=& 
\frac{G_F}{\sqrt{2}} V_{ub}V_{ud}^* a_2  f_{\pi}
(\epsilon^*\cdot p_{\pi^-}) \left[ ( m_B + m_\rho)A_1^{B\to \rho}(0)-
(m_B-m_\rho) A_2^{B\to \rho}(0) \right] \nonumber\\  \label{s12}
&=& 
\frac{G_F}{\sqrt{2}} V_{ub}V_{ud}^* a_2  f_{\pi}
(\epsilon^*\cdot p_{\pi^-}) 2 m_\rho A_0^{B\to \rho}(0).
\end{eqnarray}
In (\ref{s10}) the pole at $q^2=m_\pi^2 =m_{B_1}^2$ has canceled, as to 
be expected.
The final result (\ref{s12}) differs from (\ref{s5}) only due to the 
cancelation of the pole, so that $A_0^{B\to \rho}$ is to be evaluated at 
$q^2=0$ instead of $q^2=m_\pi^2$ in (\ref{s5}). 
It is clear that this change makes only a small difference since 
$A_0^{B\to \rho}(m_\pi^2)\simeq A_0^{B\to \rho}(0)$.

The relation between the two different calculations of the pole diagram in 
Fig.\ref{f4}b is quite analogous to the results obtained for $\bar B^0 \to 
\pi^+ \pi ^-$ in the previous section. 
We obtain analogous results for the decays  $\bar B^0 \to 
D^{*+} D_s ^-$, $D^{*+} D ^-$, $D^{*+} \pi ^-$ and $D^{*+} K ^-$. 
Of course, for some of these channels the results (\ref{s5}) and (\ref{s12})
differ by a larger amount, since in (\ref{s5}) the form factor 
$A_0^{B\to D^*}$ must be calculated for a larger $q^2$ instead of $q^2=0$ in
(\ref{s12}).

Similarly to the decay channel  $\bar B^0 \to D^{+} \pi ^-$ we have also 
contributions of the annihilation diagram for the channels  $\bar B^0 \to 
D^{+} \rho ^-$ and  $\bar B^0 \to D^{*+} \pi ^-$. 
If we evaluate them with the usual BSW approach or in the pole model with 
intermediate $0^-$ poles we obtain a contribution proportional to $A_0
^{D\to \rho}(m_B^2)$ for $\bar B^0 \to D^{+} \rho ^-$ and $A_0
^{D^* \to \pi}(m_B^2)$ for the channel $\bar B^0 \to D^{*+} \pi ^-$.
These form factors, which must be evaluated at $q^2=m_B^2$, are suppressed. 
Therefore we shall neglect the annihilation contributions as we have done 
also for the  $\bar B^0 \to D^{+} \pi ^-$ decay. 
Only when we follow ref.\cite{11}, where the pole term is canceled by the 
same factor in the numerator, we arrive at a somewhat larger result which 
can be expressed again by $A_0^{D\to \rho}(0)$ and $A_0^{D^* \to \pi}(0)$, 
respectively. 
For example, for the channel $\bar B^0 \to D^{+} \rho ^-$ the final result is:
\begin{equation}
  \label{s13}
  <D ^+ \rho ^- | {\cal H}_{eff} | \bar B^{0} >= -
\frac{G_F}{\sqrt{2}} V_{cb}V_{ud}^* a_1  f_{B}
(\epsilon^*\cdot p_{D^+}) 2 m_\rho A_0^{D\to \rho}(0).
\end{equation}
To derive this result we used the strong coupling Lagrangian for the 
coupling of the $1^+$ resonance $D_1^0$ to $D^+\rho^-$ and then introduced $
A_1^{D\to \rho}$ and $A_2^{D\to \rho}$ with (\ref{s11}). 
A similar formula as (\ref{s13}) can be derived for the annihilation term
of the decay $\bar B^0 \to D^{*+} \pi^{-}$. 

In the same way we calculate the class I decays of the charged $B^-$.
These are the channels $B^- \to D^0 D_s^{*-}$ and $D^0D^{*-}$, which have 
amplitudes proportional to $F_1^{B\to D}$, and the channels $B^- 
\to D^{*0} D_s^{-}$ and $D^{*0}D^{-}$ with amplitudes proportional to 
$A_0^{B\to D^*}$.

The class II decays with amplitudes proportional to $a_2$ are 
calculated in the same way. 
The amplitudes are either proportional to  $F_1$ or to $A_0$. 
The decays $\bar B^0\to \pi^0 J/\psi$, $\bar K^0 J/\psi$ and $B^-\to \pi^-
 J/\psi$, $ K^- J/\psi$ belong to the first group.
The second group consists of the decays $\bar B^0\to \rho^0 \eta_c$, $
\bar K^{*0} \eta_c$ and $ B^-\to \rho^- \eta_c$, $ K^{*-} \eta_c$.
Other class II decays are the channels $\bar B^0\to D^{*0} \pi^0 $,
$ D^{0} \rho^0 $ and $\bar B^0\to \rho^{0} \pi^0 $.

The class III decays with amplitudes coming from both operators $O_1$ and 
$O_2$ are the following: 
$B^- \to D^0\rho^-$, $D^0 K^{*-}$ and $B^- \to D^{*0}\pi^-$, $D^{*0}K^-$ 
originating from $b\to c$ transitions. 
The channels of class III with $b\to u$ transitions are $B^-\to \rho^0\pi^-$ 
and $\rho^-\pi^0$.

\begin{small}
\begin{table}\begin{center}
\caption{$B\to PV$ class I decay amplitudes($10^{-7}$) 
and branching ratios in two versions of pole models and experimental 
measured branching ratios or upper limits (90\% C.L.) \cite{16,20,dp}. }
\begin{tabular} {|l|cc|cc|c|}
\hline
 & \multicolumn{2}{|c|}{Pole model} & \multicolumn{2}{|c|}{BSW model} 
& Experimental \\
Channel & Amplitude &  BR &Amplitude&  BR & BR \\
\hline
$\bar B^0 \to \pi^+  \rho^-$  & 0.033  $a_2$ & $2.84\times 10^{-5}$  & 
 0.033  $a_2$ & $2.84\times 10^{-5}$ & $<8.8\times
10^{-5}$\\
$\bar B^0 \to D^+  \rho^-$  & 0.73$a_2$-0.44$a_1$ & $6.60\times 10^{-3}$  & 
 0.73  $a_2$ & $9.21\times 10^{-3}$ & $(7.8\pm 1.4)
10^{-3}$\\
$\bar B^0 \to D^+  K^{*-}$  & 0.19  $a_2$ & $4.62\times 10^{-4}$  & 
 0.19  $a_2$ & $4.62\times 10^{-4}$ & $ -$\\
$\bar B^0 \to D^+  D^{*-}$  & 0.53  $a_2$ & $3.72\times 10^{-4}$  & 
 0.53  $a_2$ & $3.72\times 10^{-4}$ & $ <1.8\times 10^{-3}$\\
$\bar B^0 \to D^+  D_s^{*-}$  & 2.88  $a_2$ & $8.90\times 10^{-3}$  & 
 2.88  $a_2$ & $8.90\times 10^{-3}$ & $<2.0\times
10^{-2}$\\
$\bar B^0 \to \pi^-  \rho^+$  & 0.016  $a_2$ & $7.01\times 10^{-6}$  & 
 0.016  $a_2$ & $7.02\times 10^{-6}$ & $<8.8\times
10^{-5}$\\
$\bar B^0 \to \pi^- D^{*+}$  & 1.02$a_2$-1.15$a_1$ & $1.37\times 10^{-3}$ &
 1.02 $a_2$ &$2.71\times 10^{-3}$ &(
$2.81\pm 0.24 ) 10^{-3}$\\
$\bar B^0 \to K^- D^{*+}$  & 0.28 $a_2$ & $1.98\times 10^{-4}$ & 
 0.28 $a_2$ &$2.01\times 10^{-4}$ &-\\
$\bar B^0 \to D^- D^{*+}$  & 0.43   $a_2$ &  $2.45\times 10^{-4}$ & 
 0.48   $a_2$ &   $2.95\times 10^{-4}$& $<1.8\times 10^{-3}$
\\
$\bar B^0 \to D_s^- D^{*+}$  & 2.20   $a_2$ &$5.75\times 10^{-3}$  &
 2.44   $a_2$ &$7.09\times 10^{-3}$ &(
$1.2\pm 0.6 ) 10^{-2}$\\
 $B^- \to  D^0 D^{*-} $  & 0.53 $a_2$ & $3.88\times 10^{-4}$ &  
  0.53 $a_2$ &$3.88\times 10^{-4}$&-\\
$B^- \to D^0 D_s^{*-} $  & 2.88   $a_2$ & $9.29\times 10^{-3}$ &
 2.88   $a_2$ & $9.29\times 10^{-3}$ &($1.2\pm 1.0 )
10^{-2}$\\
 $B^- \to  D^{*0} D^{-} $  & 0.43 $a_2$ & $2.56\times 10^{-4}$ &  
  0.475 $a_2$ &$3.07\times 10^{-4}$&-\\
$B^- \to D^{*0} D_s^{-} $  & 2.20   $a_2$ & $6.01\times 10^{-3}$ &
 2.44   $a_2$ & $7.37\times 10^{-3}$ &($10\pm 7 )
10^{-3}$\\
\hline
\end{tabular}\end{center}
\end{table}
\end{small}

We have calculated all these decays again in the two versions, (i) pole model 
without canceling of the pole of the intermediate state, which gives 
the same results as the usual BSW calculation and (ii) pole-model in the 
form as introduced in ref.\cite{11}. 
This model has a larger annihilation contribution and in the direct 
contribution, where 
the $0^-$ intermediate state is replaced by the $1^+$ state, the form factor 
is evaluated at $q^2=0$ instead of $q^2= m_\pi^2$ or $q^2= m_D^2$ as it 
appears in version (i).

\begin{small}
\begin{table}\begin{center}
\caption{$B\to PV$  class II and III decay amplitudes($10^{-7}$)
 and branching ratios in two versions of pole models and experimental 
measured branching ratios or upper limits (90\% C.L.) \cite{16,20,dp}. }
\begin{tabular} {|l|cc|cc|c|}
\hline
 & \multicolumn{2}{|c|}{Pole model} & \multicolumn{2}{|c|}{BSW model} 
&Experimental \\
Channel & Amplitude &  BR &Amplitude&  BR &  BR \\
\hline
$\bar B^0 \to \pi^0 J/\psi $  & 0.58   $a_1$ &$1.09\times 10^{-5}$ &
 0.58   $a_1$ &$1.09\times 10^{-5}$ &$<6.9\times 10^{-3}$\\
$\bar B^0 \to \bar K^0 J/\psi$  & 4.07   $a_1$ &  $4.89\times 10^{-4}$ &
 4.07   $a_1$ &  $4.89\times 10^{-4}$  & $(8.5^{+1.5}_{-1.3} )10^{-4}$
\\
$ B^- \to \pi^- J/\psi$  & 0.83   $a_1$ & $2.25\times 10^{-5}$ &
 0.83   $a_1$ & $2.25\times 10^{-5}$ &$<4.4\times 10^{-5}$ \\
$ B^- \to K^- J/\psi$  & 4.07   $a_1$ & $5.09\times 10^{-4}$ &
 4.07   $a_1$ & $5.09\times 10^{-4}$ & $(10.2\pm 1.1)10^{-4}$\\
$\bar B^0 \to \rho^0 \eta _c$  & 0.08 $a_1$ & $3.34\times 10^{-6}$ &
 0.12 $a_1$ &$7.20\times 10^{-6}$ & -\\
$\bar B^0 \to \bar K^{*0} \eta_c$  & 0.69 $a_1$ & $1.58\times 10^{-4}$ &
 0.99 $a_1$ &$3.30\times 10^{-4}$ &$- $\\
$B^- \to \rho^- \eta_c$  & 0.12 $a_1$ & $6.94\times 10^{-6}$ &
 0.17 $a_1$ &$1.49\times 10^{-5}$ &$- $\\
$B^- \to K^{*-} \eta_c$  & 0.69 $a_1$ & $1.64\times 10^{-4}$ &
 0.99 $a_1$ &$3.42\times 10^{-4}$ &$- $\\
$\bar B^0 \to \pi^0 \rho^{0}$  & 0.024 $a_1$ & $1.03\times 10^{-6}$ &
 0.024 $a_1$ &$1.03\times 10^{-6}$ &$<2.4\times 10^{-5}$\\
$\bar B^0 \to \pi^0 D^{*0}$  & 0.84 $a_1$ & $1.20\times 10^{-4}$ &
 0.84 $a_1$ &$1.20\times 10^{-4}$ &$<9.7\times 10^{-4}$\\
$\bar B^0 \to \rho^0 D^{0}$  & 0.23 $a_1$ & $6.22\times 10^{-5}$ &
 0.27 $a_1$ &$8.01\times 10^{-5}$ &$<5.5\times 10^{-4}$\\
\hline
$ B^- \to \rho^- \pi^0$  & $0.02 a_2$+0.01$a_1$ &$1.87\times 10^{-5}$  &
$0.01 a_2$+0.02$a_1$ &$1.87\times 10^{-5}$ & $<7.7 \times 10^{-5}$\\
$ B^- \to \rho^0 \pi^-$  & $0.01 a_2$+0.02$a_1$ &$8.39\times 10^{-6}$  &
 $0.01 a_2$+0.02$a_1$ & $8.39\times 10^{-6}$ & $<4.3 \times 10^{-5}$\\
$ B^- \to \rho^- D^0$  & $0.73 a_2$+0.33$a_1$ &$1.19\times 10^{-2}$  &
 $0.73 a_2$+0.38$a_1$ & $1.19\times 10^{-2}$ &
$(1.34\pm 0.18)10^{-2}$\\
$ B^- \to \pi^- D^{*0}$  & $1.02a_2$+1.19$a_1$ &$4.74\times 10^{-3}$ &
$1.02a_2$+1.19$a_1$ & $4.74\times 10^{-3}$ & $(4.34\pm 0.51) 10^{-3}$\\
$ B^- \to K^- D^{*0}$  & $0.28a_2$+0.32$a_1$ &$3.42\times 10^{-4}$ &
$0.28a_2$+0.32$a_1$ & $3.45\times 10^{-4}$ & - \\
$ B^- \to K^{*-} D^0$  & $0.19a_2$+$0.10a_1$ &  $6.15\times 10^{-4}$ &
$0.19a_2$+$0.11a_1$ & $6.35\times 10^{-4}$ &-
\\
\hline
\end{tabular}\end{center}
\end{table}
\end{small}

The results for the class I decays are collected in Table~8 and for the 
class II and class III decays in Table~9. 
In these two tables, the amplitudes are defined without the factor of $(
\epsilon \cdot p_\pi)$ in (\ref{s7}).
In Table~8, four of the decay channels can be compared to experimental data.
In all four channels the agreement is rather good. 
Unfortunately the experimental errors for the $D^{(*)} D_s^{(*)}$ final 
states are rather large. 
For $\bar B^0 \to \pi^-D^{*+}$ the error is smaller and only the prediction of 
the BSW pole model agrees with the experimental number. 
 The prediction of the pole model with canceling residues is a factor
of two smaller than the experimental branching ratio. 
This is due to the additional annihilation contribution in this model 
which is too large when compared with the experimental branching ratio. 
Since in this model the off-shell extrapolation of the form factor to the 
$q^2=m_B^2$ is neglected the discrepancy with the data is not unexpected.
Actually the disagreement of this pole model in the case 
of $\bar B^0 \to \pi ^- D^{*+}$ is a strong argument against this version
 of pole model. Of course, this depends on the values assumed for $a_1$ 
and $a_2$, i.e. our assumption $N=2$ also in this pole model. 
If $a_1$ would be decreased as it happens for example, with $N=3$, i.e. 
in pure factorization with colour-octet terms omitted, the annihilation 
term can be suppressed. 
However, as we shall see later, this leads to strong disagreement in other 
channels, where the two pole model versions have identical results.
 From this comparison it is clear that the extra form factor in the 
annihilation diagram must be taken into account also in this pole model. 
This makes these contributions negligible also.
This decay gives us a good value for $A_0^{B\to D^*}(0)$.
We notice also that the prediction for  $\bar B^0 \to D_s^- D^{*+}$ agrees in 
the BSW-pole-model slightly better with the experimental number than in 
the other pole model.
In this case the different results in the two models come only from 
the usual W emission graph.

As already mentioned above, the two pole models give identical results for 
the decay channels $\bar B^0 \to D^+K^{*-}$,  $D^+D^{*-}$, $D^+D_s^{*-}$
and $\pi^+\rho^{-}$.
Differences occur for the decays  $\bar B^0 \to D^-D^{*+}$ and
 $D_s^-D^{*+}$.

In Table~9, our results for the branching ratios of the decays $\bar B^0 \to
\bar K^0 J/\psi$ and  $B^- \to K^- J/\psi$ are somewhat smaller than the 
measured branching ratios. 
The factor r needed to achieve agreement is $r=1.7\pm 0.3$ for $\bar B^0 \to
\bar K^0 J/\psi$ and $r=2.0\pm 0.2$ for $B^- \to K^- J/\psi$, respectively.
Since the Wilson coefficient $a_1$ is already constrained by the class III 
decays the disagreement means either a $(30-40)\%$ larger value for 
$F_1^{B\to K}(0)$ or a breakdown of the factorization assumption \cite{17}.
It is clear that other values for $a_1$, which for example for $N=3$ in 
the pure factorization approach with colour octet terms neglected are obtained,
would give very bad results for both pole models. 
For some of the other decays of class II upper limits for the branching 
ratios exist which are all obeyed by the theoretical results.

The prominent class III decays are  $B^- \to \rho^- D^0$ and $B^- \to \pi^- 
D^{*0}$. The predictions agree quite well with the experimental branching
 ratios. 
It is clear that this agreement, given the matrix elements of BSW, is
 only possible, if $a_1$ is positive, as was already found in ref.\cite{16}.
This statement is again independent of the form of pole model applied 
in the calculation.
Concerning matrix elements these two decays depend on $F_1^{B\to D}$, 
$F_1^{B\to \pi}$, $A_0^{B\to \rho}$ and $A_0^{B\to D^*}$.

Next we consider the $B\to VV$ decays. 
For these decays the two pole-model versions give identical results
 which are the same as in the quark diagram approach.
So, we need not write down all the formulas. We do this only for one 
channel $\bar B^0 \to \rho^+\rho^-$ to fix our notation. 
The axial part of the current matrix  element was already written 
 in (\ref{s3}). 
In addition we need also the matrix element of the vector current which we 
write as usual
\begin{equation}
  \label{36}
  <\rho^+| \bar u \gamma_\mu b | \bar B^0> =\epsilon _{\mu\nu\alpha\beta}
\epsilon^{\nu*} p_B^\alpha p_\rho^\beta \frac{2 V(q^2)}{(m_B+m_\rho)}.
\end{equation}
With this, (\ref{9}) and (\ref{s3}) we obtain for the weak transition 
matrix element 
\begin{eqnarray}
  <\rho^+\rho^- | {\cal H}_{eff} | \bar B^0> &=&
\frac{G_F}{\sqrt{2}} V_{ub}V_{ud}^* (-i f_\rho m_\rho) \left \{
\epsilon^* (\rho^+) \epsilon ^*(\rho^-) (m_B +m_\rho) A_1^{B\to \rho }
(m_\rho^2)\right.\nonumber\\
&-&(\epsilon^* (\rho^+)\cdot p_B)( \epsilon^*(\rho^-)\cdot p_B) \frac {
 2 A_2^{B\to \rho }(m_\rho^2)}{ (m_B +m_\rho)}\nonumber\\
&-&i\left.\epsilon _{\mu\nu\alpha\beta} \epsilon^{\mu *}(\rho^-) 
\epsilon^{\nu *}
(\rho^+) p_B^\alpha p_{\rho^+}^\beta \frac{2 V^{B\to \rho}(m_\rho^2)}
{(m_B+m_\rho)}\right \}.  \label{37}
\end{eqnarray}
So, the transition matrix element is a linear combination of the form 
factor $A_1$, $A_2$ and $V$ which determine the transition in the s, d 
and p waves respectively \cite{x}.

\begin{small}
\begin{table}\begin{center}
\caption{$B\to VV$  decay amplitudes($10^{-8}$) 
and branching ratios in pole model and experimental 
measured branching ratios or upper limits (90\% C.L.) \cite{16,20,dp}. }
\begin{tabular} {|l|ccc|c|c|}
\hline
 &\multicolumn{3}{|c|}{ Amplitudes} &  model
& Experimental \\
Channel &  V($10^{-8}$) & A1($10^{-8}$) & A2($10^{-8}$) &  BR & BR \\
\hline
$\bar B^0 \to \rho^+  \rho^-$  & 1.94  $a_2$ & 
 0.84  $a_2$ & 1.67  $a_2$ &  $2.40\times 10^{-5}$ & $<2.2\times
10^{-3}$\\
$\bar B^0 \to D^{*+}  \rho^-$  & 54.1  $a_2$
& 25.0  $a_2$ &52.6 $a_2$
& $8.70\times 10^{-3}$ & $(7.3\pm 1.5)10^{-3}$\\
$\bar B^0 \to D^{*+}  K^{*-}$  & 14.3 $a_2$ & 
 6.59  $a_2$ &  13.9  $a_2$& $4.68\times 10^{-4}$ & $ -$\\
$\bar B^0 \to D^{*+}  D^{*-}$  &  39.8  $a_2$  & 
 18.1  $a_2$ & 38.2  $a_2$  &  $9.28\times 10^{-4}$ & 
$ (5.3 ^{+7.2}_{-3.8})10^{-4}$\\
$\bar B^0 \to D^{*+}  D_s^{*-}$  & 215  $a_2$  & 
 97.8  $a_2$ &206  $a_2$  & $2.46\times 10^{-2}$ & $(1.9\pm 1.2)
10^{-2}$\\
 $B^- \to  D^{*0} D^{*-} $   & 39.8  $a_2$  & 
  18.1 $a_2$ &38.2  $a_2$   &$8.87\times 10^{-4}$&-\\
$B^- \to  D^{*0}D_s^{*-} $   & 215  $a_2$  &
97.8   $a_2$ &206  $a_2$ & $2.56\times 10^{-2}$ &($2.3\pm 1.4 )
10^{-2}$\\
\hline
$\bar B^0 \to \rho^0 J/\psi $  &34.8  $a_1$ &
 14.9   $a_1$ &29.7  $a_1$ &$5.16\times 10^{-5}$ &$-$\\
$\bar B^0 \to \bar K^{*0} J/\psi$   &245  $a_1$ &
101   $a_1$  &204 $a_1$ &  $2.03\times 10^{-3}$  & $(1.32 \pm 0.24 )10^{-3}$
\\
$ B^- \to \rho^- J/\psi$  &49.3  $a_1$  &
 21.0   $a_1$ &42.0  $a_1$ & $1.07\times 10^{-4}$ &$-$ \\
$ B^- \to K^{*-} J/\psi$   &245  $a_1$ &
 101   $a_1$  &204  $a_1$& $2.11\times 10^{-3}$ & $(1.41\pm 0.33)10^{-3}$\\
$\bar B^0 \to \rho^0 \rho^{0}$   & 0.97  $a_1$ &
 0.42 $a_1$ &0.84  $a_1$ & $7.74\times 10^{-7}$ &$<2.8\times 10^{-4}$\\
$\bar B^0 \to \rho^0 D^{*0}$  &50.3  $a_1$ &
 21.6 $a_1$ &43.1  $a_1$ & $2.40\times 10^{-4}$ &$<1.17\times 10^{-3}$\\
\hline
$ B^- \to \rho^- \rho^0$  &1.37  $(a_2+a_1)$ &  0.59 $(a_2+a_1)$
&1.18 $(a_2+a_1)$ &$1.95\times 10^{-5}$ & $<1.0 \times 10^{-3}$\\
$ B^- \to \rho^- D^{*0}$   &54.1$a_2$+71.1$a_1$  &
 $24.9 a_2$+30.5$a_1$ &52.6$a_2$+61.0$a_1$& $1.37\times 10^{-2}$ &
$(1.55\pm 0.31)10^{-2}$\\
$ B^- \to K^{*-} D^{*0}$  &14.3$a_2$+18.4$a_1$ &
$6.59a_2$+7.95$a_1$ & 13.9$a_2$+16.1$a_1$ &$7.50\times 10^{-4}$ & - \\
\hline
\end{tabular}\end{center}
\end{table}
\end{small}

We have calculated the branching ratios for various decays using the form
factor values given in Table~3 together with the decay constants in
Table~2 and the pole masses from Table~1.
The results are given in Table~10 and are compared to the experimental 
branching ratios (last column in Table~10). 
The decay channels in Table~10 are ordered according to class I, II and III 
as in the previous tables. 
The calculated branching ratios agree well with the experimental data, in
particular for the decays $\bar B^0 \to D^{*+}\rho^-$, $ D^{*+} D_s^{*-}$
 and $B^-\to D^{*0}D_s^{*-}$ which are class I decays and $B^-\to \rho^- 
D^{*0}$ which is a class III decay. 
The class II decays $\bar B^0 \to K^{*0} J/\psi $ and $ B^- \to K^{*-} J/\psi $
come out somewhat larger than the measured branching ratios. 
This could be changed by adjusting the $B\to K^*$ form factor by a small 
amount or may be a  sign of the breakdown of the factorization assumption for 
class II decays \cite{17}. 

In Table~10 we have collected also the amplitudes proportional to the form
 factors $V^{B\to \rho}$, $A_1^{B\to \rho}$ and    $A_2^{B\to \rho}$.
The numbers in the first, second and third column of Table~10 are defined
 in such a way that the polarization dependent factors 
$[-\epsilon _{\mu\nu\alpha\beta} \epsilon^{\mu*}(\rho^-) \epsilon^{\nu*}
(\rho^+) p_B^\alpha p_{\rho^+}^\beta /(m_B+m_\rho)^2]$  in the term
proportional to V in (\ref{37}),  $[\epsilon^*(\rho^+)
\epsilon^*(\rho^-)]$ in the term proportional to $A_1$ in (\ref{37})
and $[ (\epsilon ^* (\rho^+)\cdot p_B)( \epsilon ^*(\rho^-)\cdot p_B) /
(m_B +m_\rho)^2]$ in the term proportional to $A_2$ in (37) are not included. 
Except for these factors the amplitudes are of the same order of magnitude 
as to be expected from the values of the form factors in Table~3. 
The results in Table~8, 9 and 10 depend on the parameters $A_1 ^{B\to \rho}$, 
$A_2 ^{B\to \rho}$, $V ^{B\to \rho}$ and $A_1 ^{B\to D^{*}}$, 
$A_2 ^{B\to D^*}$, $V ^{B\to D^*}$, respectively. The BSW values can be 
compared to results from other sources, i.e. either other quark models,
 QCD sum rules or lattice gauge theory. 
Values for $A_1 ^{B\to \rho}$, $A_2 ^{B\to \rho}$ and $V ^{B\to \rho}$,
reported in the literature, are collected in Table~11.
Except for a few cases, the form factors are similar to the BSW values. 
 From lattice gauge theory calculation the following values for 
$A_1 ^{B\to \rho}(0)$ are reported: $0.22\pm 0.05 $ \cite{t12}, 
$0.24\pm 0.12 $ \cite{t13} and $0.27^{+0.07}_{-0.04}$ \cite{s11}, for 
a recent review see \cite{s16}. 
A similar collection for $A_1 ^{B\to D^{*}}$, $A_2 ^{B\to D^*}$, 
$V ^{B\to D^*}$ is found in Table~12. 
$A_0^{B\to \rho}$ and $A_0^{B\to D^*}$ can be calculated with the help of
(\ref{s4}) from $A_1$ and $A_2$, respectively, given in Table~11 and Table~12.

It is clear that the good agreement of the calculated partial decay rates 
with the experimental data in the case of the $B\to PV$ and $B\to VV$ decays 
 serves only as a consistency check of the form factors $A_1$, $A_2$ and $V$
 for $B\to \rho$ and $B\to D^*$. 
In order to obtain complete information on these form factors many more 
measurements are needed than just the branching ratios. 
Additional information can come from decay angular distributions of 
subsequent decays of the vector mesons. 
This has been studied in previous work \cite{x,y} and compared to 
experimental data in \cite{16}.

\begin{table}[htbp]
  \begin{center}   
 \caption{Form factors $A_1^{B\to \rho}$, $A_2^{B\to \rho}$, $V^{B\to \rho}$ 
at $q^2=0$ in different models. }
    \begin{tabular}{|cc|c|c|c|}
\hline
Reference & method &  $A_1^{B\to \rho}(0)$ & $A_2^{B\to \rho}(0)$&
 $V^{B\to \rho}(0)$\\
\hline
\cite{57} & $c$ & $0.24\pm 0.04$ & -- & $0.28\pm 0.08$\\
\cite{t7} & $c$ & $0.5\pm 0.1$ & $0.4\pm 0.2$ & $0.6\pm 0.2$\\
\cite{t9} & $c$ & $0.35\pm 0.16$ & $0.42\pm 0.12$ & $0.47\pm 0.14$\\
\cite{64} & $c$ & $0.27\pm 0.05$ & $0.28\pm 0.05$ & $0.35\pm 0.07$\\
\cite{7} & $a$ & $0.283$ & $0.283$ & $0.329$\\
\cite{t15} & $a$ & $0.27$ & $0.30$ & $0.29$\\
\cite{65} & $a$ & $0.26$ & $0.24$ & $0.35$\\
\cite{t5} & $b$ & $0.21$ & $0.20$ & $1.04$\\
\hline
    \end{tabular}
    \label{tab:11}
  \end{center}
\end{table}

We are now in the position to calculate other strong coupling 
constants of the B mesons, which enter into the pole-model formulation 
of non-leptonic weak decays, on the basis of the information following 
from the form factors $A_1$, $A_2$ and $V$ for $B\to \rho$, $B\to K^*$ 
and $B\to D^*$. 
As a reference we employ the BSW results for these form factors as given
 in Table~3. 
It is clear that the calculation can be repeated for any other choice 
as collected in Table~11 and 12. 
Let us start with $g_{BB\rho}$. 
This follows from (\ref{s6}) with the result $g_{BB\rho}=1.54$. 
In the same way we obtain $g_{BB_sK^*}=1.81$ and $g_{BB_cD^*}=3.48$. 
We observe a large splitting of the coupling between two pseudoscalar 
B mesons (BB, $BB_s$, $BB_c$) and the vector mesons $\rho$, $K^*$ and 
$D^*$, respectively. 
This is mainly due to the mass factor of the vector meson in (\ref{s6}).

\begin{table}
  \begin{center}   
 \caption{Form factors $A_1^{B\to D^*}$, $A_2^{B\to D^*}$, $V^{B\to D^*}$ 
at $q^2=0$ in different models. }
    \begin{tabular}{|cc|c|c|c|}
\hline
Reference & method &  $A_1^{B\to D^*}(0)$ & $A_2^{B\to D^*}(0)$&
 $V^{B\to D^*}(0)$\\
\hline
\cite{t9} & $c$ & $0.46\pm 0.02$ & $0.53\pm 0.09$ & $0.58\pm 0.03$\\
\cite{7} & $a$ & $0.651$ & $0.686$ & $0.705$\\
\cite{t15} & $a$ & $0.62$ & $0.61$ & $0.63$\\
\cite{23} & $f$ & $0.622$ & $0.778$ & $0.747$\\
\hline
    \end{tabular}
    \label{tab:12}
  \end{center}
\end{table}

The other coupling constants which follow from $A_1(0)$ and $A_2(0)$ with 
the relations (\ref{s11}) are the coupling constants of axial vector mesons 
$B_1$, $B_{s1}$ and $B_{c1}$ with $B\rho$, $BK^*$ and $BD^*$, respectively. 
 From $V(0)$ we obtain the coupling constants of the vector mesons $B^*$, 
$B_s^*$ and $B_c^*$ with $B\rho$, $BK^*$ and $BD^*$.
The formula for this coupling in the case of $B\rho$ in analogous to the 
second formula in (\ref{s11})
\begin{equation}
  \label{38}
  g= \frac{m_{B^*}^2}{(m_B+m_\rho)f_{B^*}} V^{B\to \rho}(0).
\end{equation}
The result of the coupling of the axial vector mesons is:
$$\begin{array}{ll}
  g_s (BB_1 \rho) = 10.7, &  g_d (BB_1 \rho) = 8.4,\\
 g_s (BB_{s1} K^*) = 11.2, &  g_d (BB_{s1} K^*) =10.3,\\
 g_s (BB_{c1} D^*) = 14.8, &  g_d (BB_{c1} D^*) =13.3.
\end{array}
$$
The corresponding vector meson coupling constants are:
\begin{eqnarray*}
  g(BB^* \rho) = 9.64, &  g(BB_s^* K^*) = 9.67, & g(BB_c^*D^*) =12.1.
\end{eqnarray*}
We observe that all coupling constants have similar values, of the order 
of 10.
The splitting of the couplings to $B\rho$, $BK^*$ and $BD^*$ is somewhat 
larger than obtained for the corresponding VPP couplings to $B\pi$, BK 
and $BD$ in section 2. 
The coupling constant $g(BB^*\rho)$ can be compared with a QCD sum rule 
computation of this coupling \cite{z}. 
Using the values for $f_B$ and $f_{B^*}$ as in our Table~2 these authors 
obtain $g(BB^*\rho)\simeq 12$ in reasonable agreement with our values above.

\section{Summary and Conclusions}

We have calculated the branching ratios for the decays of neutral and 
charged B mesons into PP, PV and VV mesons using two versions of 
pole-dominance model in addition with a factorization assumption. 
The first pole model gives the same result as the extensively studied 
factorization model \cite{7} in terms of current matrix elements, if these 
current matrix elements are approximated by single poles. 
The second pole model due to Bedaque et al.\cite{11} employs a different 
off-mass-shell extrapolation of the residues of the single-pole dominance 
approximation which leads, besides other differences, to an enhancement 
of the annihilation diagram contributions. 
  From the derivation we see that the two versions of pole models are related.
The pole model of ref.\cite{11} involves a different way of extrapolation 
away from the pole position which has a particular large effect for some 
decays where an annihilation contribution is present. In the particular 
decay $\bar B^0 \to \pi^- D^{*+}$ the model of ref.\cite{11} disagrees with 
the experimental branching ratios due to a large annihilation contribution. 
In the cases where annihilation diagrams do not contribute the final 
results of the two models do not differ very much. 
For the cases $B\to VV$ and most of the $B\to PV$ decays considered 
in this work they even give identical results.

The coupling constants in the pole model are related to the usual form 
factors of the current matrix elements at $ q^2=0$. 
For an overview we have compared the form factors $F_1^{B\to \pi}$ and 
$F_1^{B\to D}$ and the form factors for the transitions $B\to \rho$ and 
$B\to D^*$ calculated with different methods and found reasonable agreement. 
Using information for the pseudoscalar and vector meson decay constants 
we deduced the various strong couplings of the vector mesons $B^*$, $B_s^*$
 and $B_c^*$ with the $B\pi$, $BK$ and $BD$ system, respectively.
We found rather similar values for these three couplings showing that the 
splitting of the form factors  $F_1^{B\to \pi}$,  $F_1^{B\to K}$ and 
$F_1^{B\to D}$
is too a large extent related to the splitting of $f_{B^*}$, $f_{B_s^*}$ 
and $f_{B_c^*}$. 
The same pattern evolves for the couplings of the axial-vector mesons 
$B_1$ and the vector mesons $B^*$ to the $B\rho$, $BK^*$ and $BD^*$ system.
It would be interesting to know whether the relations for the strong 
coupling constants found in this work could be obtained from QCD sum 
rule or lattice calculations.

\section*{Acknowledgment}

One of us (G.K.) thanks W.F. Palmer for collaboration in the early stages
 of this work.

\end{document}